\documentclass[eqno,twoside,12pt]{article}
\usepackage[dvips]{graphicx}
\usepackage{epsfig}
\usepackage{amsfonts}
\usepackage{amssymb}

\def\lewz#1{\leqno(#1)}
\newcommand{\ak}{\hspace{20pt}}

\newcommand{\ms}{\medskip}
\newcommand{\bs}{\bigskip}

\newcommand{\si}{\sigma}

\newcommand{\ps}{\rightarrow}

\newcommand{\tg}{{\rm tg}}

\begin{document}

\parindent=0pt

\baselineskip=20pt

\vspace*{3cm}
\begin{center}
{\LARGE\bf Statistical description of magnetic\\[3mm]
domains in the Ising model}

\bs
\bs
\bs
K. Lukierska-Walasek

\ms
\em Institute of Physics

University of Zielona G\'ora

ul. Z. Szafrana 4a, 65--516 Zielona G\'ora, Poland

\bs
\bs
\em K. Topolski

\ms
\em Institute of Mathematics

Wroc{\l}aw University

pl. Grunwaldzki 2/4, 50--384 Wroc{\l}aw, Poland

\end{center}

\newpage
\pagestyle{plain}

{\small {\bf Abstract:}

We use the Mandelbrot-Zipfs power law for the description of the inhomogenity
of the spin system. We describe the statistical distributions of
the domain's masses in the Ising model near the phase transition induced by the
temperature. The statistical distribution near the critical point appears
to be of the Pareto type.
}

\vspace{0.8cm}
\ak We study in this paper for the description of the phase transition the
Ising model [1]. This model is one of the simplest models which can be used to
describe the phase transition in ferromagnet. Besides this model is known due to
several applications: first of all the percolation [2], then trading activity [3],
sociophysics [4] and others. A common feature in these problems is
the presence of two choices of the variable $\,S=\pm 1$. However all the
results one can generalize to the model containing many values of $\,S\,$ for
instance to the Potts model.

\ak The Hamiltonian for the simplest Ising model is:

\vspace{-3mm}
$$
H=-\frac{1}{2}\sum_{ij} J_{ij} \;S_i^z \;S_j^z\lewz{1}
$$

\ms
with the sum over all neighbour pairs (z-$th$ component) of spins. Usually it is
assumed that the crystal lattice of ferromagnet is regular and in each site of
a lattice the spin is localized with the value $\,S^z=1\,$ or $\,S^z=-1$.
Further

$$
J_{ij}=\left\{
\begin{array}{l}
J \quad \mbox{if \ $i,j$ are neighbour pairs of spins}\\[2mm]
0 \quad \mbox{in oposite case}.
\end{array}\right\}
$$

\bs
Two spins $\,i\,$ and $\,j\,$ interact with each other by an energy $-J\,$
$S_i^z\,$ $S_k^z\,$ with $-J\,$ if both spins are parallel and $+J\,$ if
they are oposite to each other. The energy needed for fliping of one spin is
$\,2J$.

\ak For the simulation in this model we will use the Monte Carlo method with
the Swendsen-Wang cluster algorithm [5].

\ak In this algorithm clusters of spins are created by introducing bonds between
neighboring spins with probability $\,P(S^z_i,S^z_j)=1-\exp(-\frac{\Delta E}
{k_B T}),\,$
where $\,k_B\,$ is Boltzmann constant, $\,\Delta E\,$
is the energy difference needed to transform a pair of equal spins to a pair
of opposite spins.
The probability is zero if spins are the same.
All such clusters are generated and then updated  by choosing a random new
spin value for each cluster and assigning it to all spins in this cluster.
The probability is zero if spins are the same.
Starting with the simmulation having the random distribution
of half of the spins up and half down and using Swendsen-Wang
algorithm with low
temperature one sees growing domains, in which spins are parallel.
We have two kind of domains: with spins up and with spins down. At last
at the temperature Curie $\,T=T_c\,$ $\,\left( T_c=\frac{2J}{k_B\,
ln(1+\sqrt{2})}\right)\,$
there appears an infinite domain in the limit $\,L\ps\infty\,$ where $\,L\,$ denotes
linear size of the system with one of the spin states being chosen.

\ak The phase transition appears in the critical point $\,T=T_c.\,$
The difference $\,M\,$ between a number of spins up and down is proportional
to the magnetisation and near critical point vanishes as $\,(T-T_c)^\beta$,
where for dimension $\,d=2$, $\,\beta=\frac{1}{8}$. The correlation length
$\,\xi\sim|T-T_c|^\nu$. The magnetisation is proportional to $\,\xi^{-\beta/\nu}$.
In a finite system in critical temperature $T_c$ one can replace $\,\xi$ by $L$,
hence $\,M\sim L^{d-\beta/\nu}=L^D\,$ with the fractal dimension
$\,D=d-\beta/\nu \ \ (d\leq 4)$.

\ak The simulation data were collected on square lattices of linear size
$\,L=500\,$ and 1000. A total of 10000 Monte Carlo (MC) time steps were used
for equilibration. The value of MC time steps required for equilibration have
been estimated from the energy time series, which is a common practice for
cluster algorithms.

\ak The main goal of this paper is the statistical description of the simple
magnetic system when we approach the critical point of phase transition
induced by the temperature.

\ak For this purpose we will consider Mandelbrot-Zipf's power law [6]:
$$
x=k^{-\frac{1}{\mu}}
$$

In our case $\,x\,$ is the number of spins up or down in the domain
(domain mass), $\,k\,$ denotes the rank order of the domain mass $\,x$.
(The greatest cluster has rank 1, smaller rank 2 and so on.)

\ak Bouchaud [5] pointed out the strong correlation between Mandelbrot-Zipf's power
law and the inhomogenity of the system: the slope (in log-log) of the straight
line is determined by $\,-\frac{1}{\mu}\,$ and characterize the inhomogenity
of the physical structure of the system $\,(-\frac{1}{\mu}=\tg\,\alpha,\,$ where
$\,\alpha\,$ the angle of the slope). The inhomogenity of the system means that
its structures become fractal and more hierarchical.

\ak In our considerations we shall concentrate on the sequence of random
variables $\,x\,$ called here the $\,[\mu]$-variables. These variables are
distributed according to the distribution of the appearance of a cluster with 
mass $\,x,\,$ and probability $\,\rho(x)\,$ which dacays as
$\,\frac{x^n_0}{x^{1+\mu}}$, 
where $\,x_0\,$ is the typical scale.
The index $\,\mu\,$ appearing in the tail of distribution $\,\rho(x)\,$
is a critical exponent.
The main property of $\,[\mu]\,$ variable is that all its moments
$\,m_q=<x^q>\,$ with $\,q\geq\mu\,$ are infinite.

\ak We are going to connect the statistics of domain masses with the process
of approaching the critical point. When we start to advance from the
paramagnetic phase to the critical point $\,(T\ps T_c)\,$ as we see on Fig (1)

\bs
\centerline{Insert Fig (1)}

\bs
\centerline{Fig 1. Log-log distribution of the domain's masses $\,x\,$ versus
the rank order}

\hspace{1.7cm}index $k$ \ $(L=1000)$.

\begin{center}
\begin{tabular}{ll}
1/T = 0.25: & ln(ln(k)) = 4,01287428 - 0,148950538*ln(k)\\
1/T = 0.35: & ln(ln(k)) = 5,55609713 - 0,21268554*ln(k)\\
1/T = 0.4:  & ln(ln(k)) = 6,89830264 - 0,261512561*ln(k)\\
1/T = 0.44068: & ln(ln(k)) = 10,4914484 - 1,014818*ln(k)
\end{tabular}
\end{center}

\bs
The angle between strait lines representing (in log-log) the
Mandelbrot-Zipf's inverse power law and the rank axis increase when $\,\mu>1$
and $T\ps T_c$.

\ak For that case we observe the growing domains, their structure become more
fractal (loss of an oval) and more hierarchical - the inhomogenity of the system
increase. The distribution of magnetisation of a whole system has the usual
Gauss form.
At high temperature correlations between spins are short ranged in the whole
high temperature region $\,M\sim\sqrt{L^d}=L\,$ for $\,d=2$, see Fig. 2.

\bs
\centerline{Insert Fig (2)}

\bs
\centerline{Fig 2. Test $\,M\sim L\,$ for $T=5\,T_c$.}

\bs
At last at $\,T=T_c\,$ and $\,\mu\approx 1$, when one domain covers
the whole lattice and the other spin orientation is restricted to the small
clusters or isolated single spins within the domain. In this case the
distribution of the total magnetisation is centered at $\,mL^d\,$ or
$\,-mL^d,\,$ where $\,m\,$ is the remnant magnetisation and it also has
Gaussian form, as it is well known.

\ak In the case when $\,\mu\approx 1\,$ $(T=T_c)\,$ from fig (1) we see the
strait line representing (when $\,\alpha=\alpha_c$, $\,\tg\,\alpha_c=
\frac{1}{\mu_c}\approx 1$) the Mandelbrot law describing the phase transition.
The highest point denotes in the critical temperature the domain
which covers the whole lattice.

\bs
Fig (3) represents the histogram of the domain masses at critical point
-- the probability of the appearence of the cluster with mass $\,x.\,$

\bs
\centerline{Insert Fig (3)}

\bs
\centerline{Fig 3. Histogram of the domain's masses at critical point
$T=T_c$ $(L=500)$}

\bs
The distribution $\,\rho(x)\,$ of domain mass is like has
Pareto tail $\,\frac{x_0^\mu}{x^{1+\mu}}$, where
$\,x_0\,$ denotes a typical scale, with $\,\mu\approx 1$.

\ak When $\,\mu>1\,$ distribution $\,\rho(x)\,$ is without power-law tail, which
becomes truncated, see Fig 4.

\bs
\centerline{Insert Fig (4)}

\bs
\centerline{Fig 4. The histogram of the domain's masses
($\beta=\frac{1}{kT}=0,3$; \ $L=500)$}

\bs
This results are in the agreement with standard percolation theory [2]
and the paper of Janke and Schakel [6] because the distribution of domains
with the mass $\,x\,$ takes a general form
$$
\varrho_x\sim x^{-\tau} \exp (-\Theta x)
$$

\ms
where $\,\tau\,$ is the entropy factor $\,(\tau=\frac{d}{D}+1)$,
$\,\Theta\sim(T-T_c)^{\frac{1}{\si}}$, if $\,T\ps T_c$. When $\,T=T_c$,
we get $\,\Theta=0\,$ and we have $\,\varrho_x\sim x^{-\tau}$.

\ak The statistical description of discontinuous metal films on dielectric
substrates were analysed on the basis of experiments by Dobierzewska-Mozrzymas
at all [8] and the distributions of local fields
intensities in metal dielectric system was investigated by Liberman
{\it et al.} [9]. Unfortunately the analogous experiments on ferromagnetics
are not known.

\ms
\ak The main result of our paper is to show the conection beetwen the
\linebreak
Mandelbrot-Zipf's law and the statistics of the domain masses in the Ising
model. The domain masses in the
Ising model fulfil the Mandelbrot-Zipf's inverse power law and when we approach
the phase transition in this model the distribution of the domain masses
appears to have the Pareto tail.
This model is a such one which represents the system in which in the critical
point the lenght scale diverges and leaves the system in self similar state.
That feature denotes a fined-tuned criticality which should be contrasted with
self-organized criticality. In such a case the system spontaneously evolves
towards scale-invariant states and one raise the problem raises problem of the
universality of the renormalized coupling constant at critical point.
This conclusion is in agreement with the paper of Hilfer [10].

\bs
\centerline{\bf Acknowledgments}

\ms
The authors would like to express their gratitude to prof.
E. Dobierzewska-Mozrzymas and dr P. Biega\'nski and dr J. Cis{\l}o for the interest
in this paper and the valuable discussions.

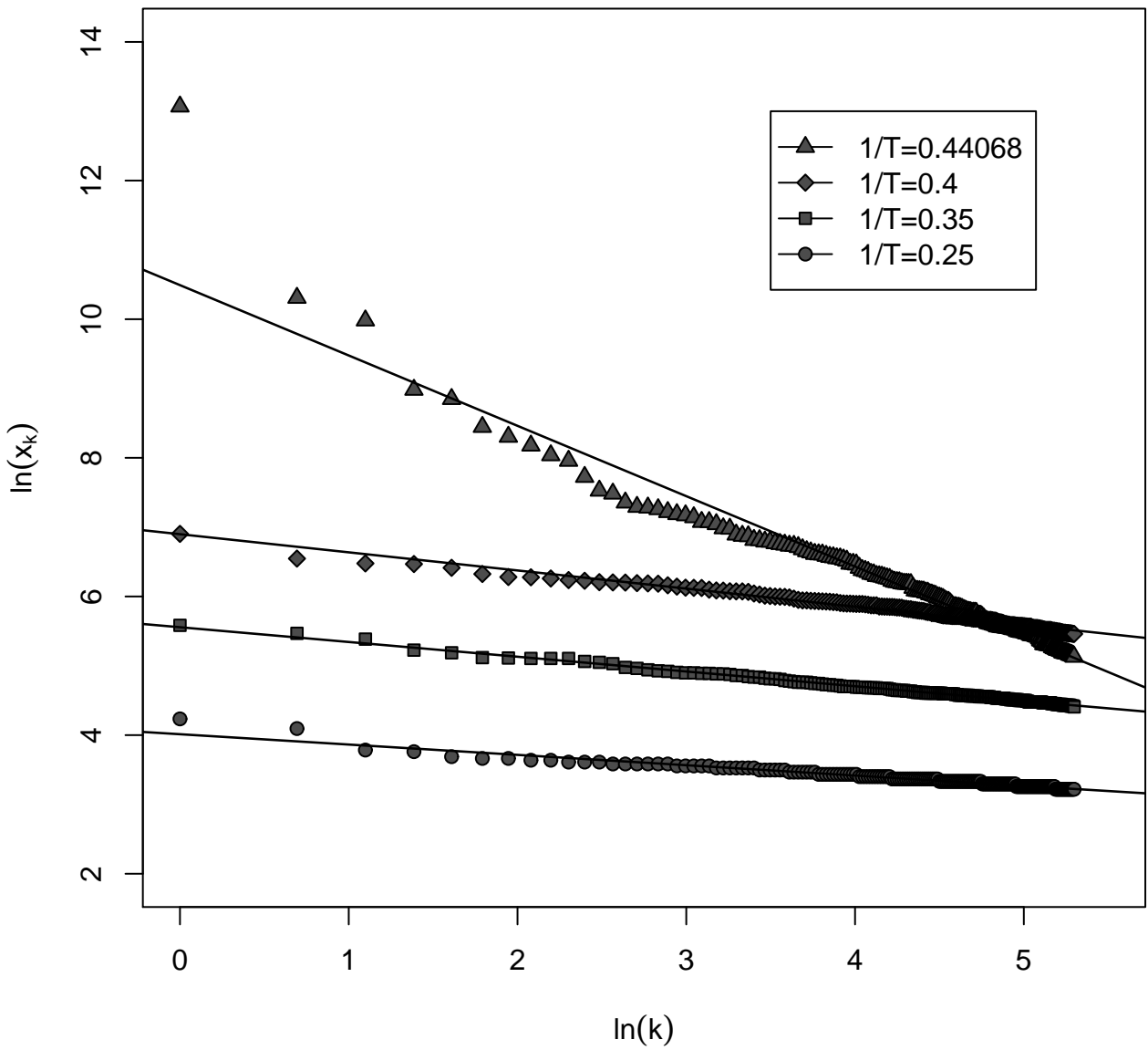
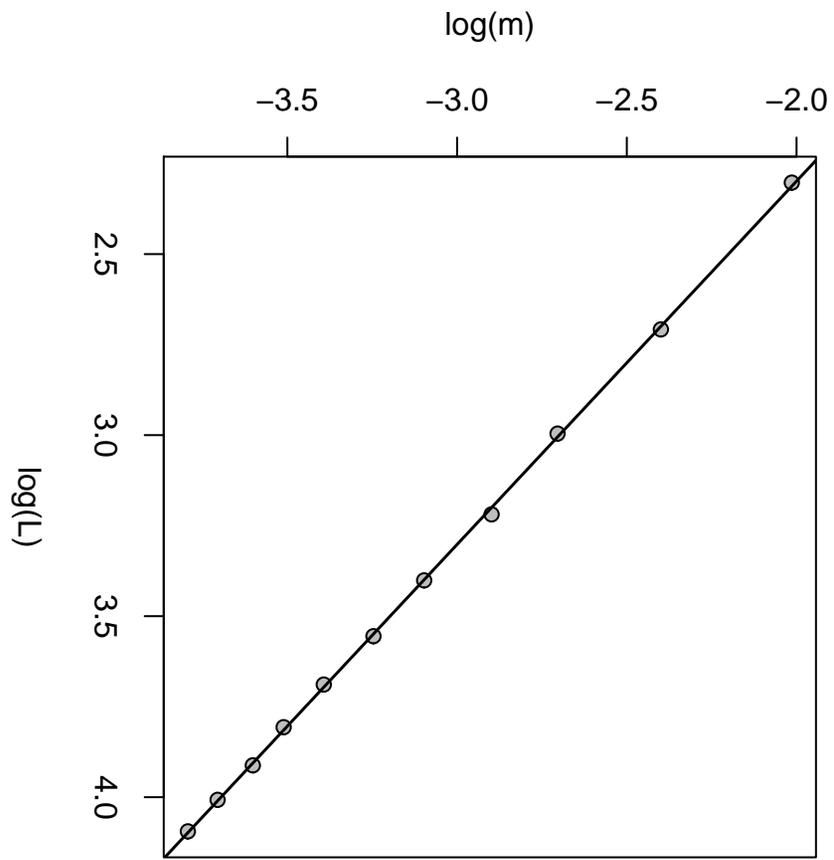
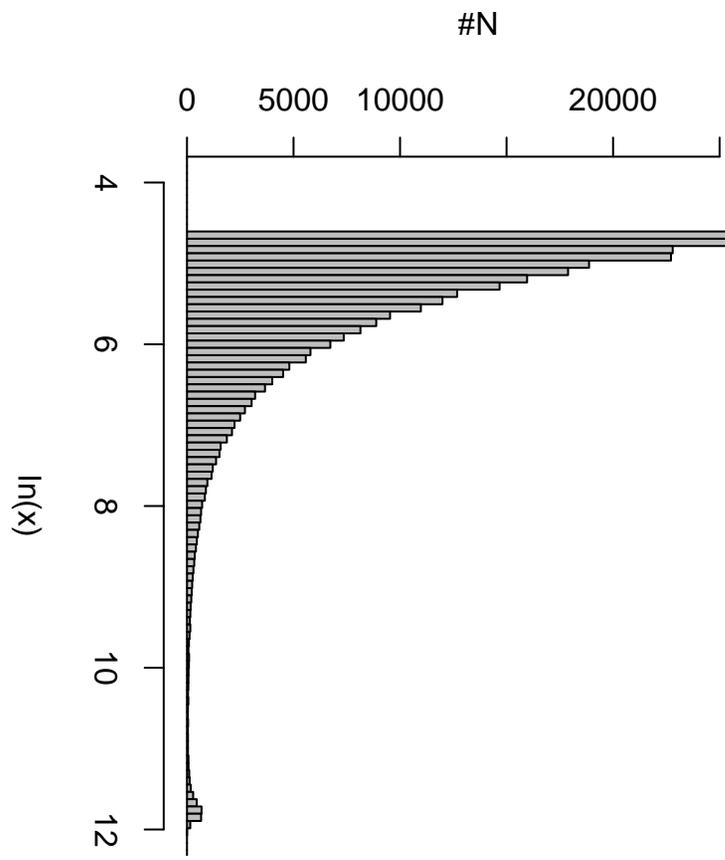
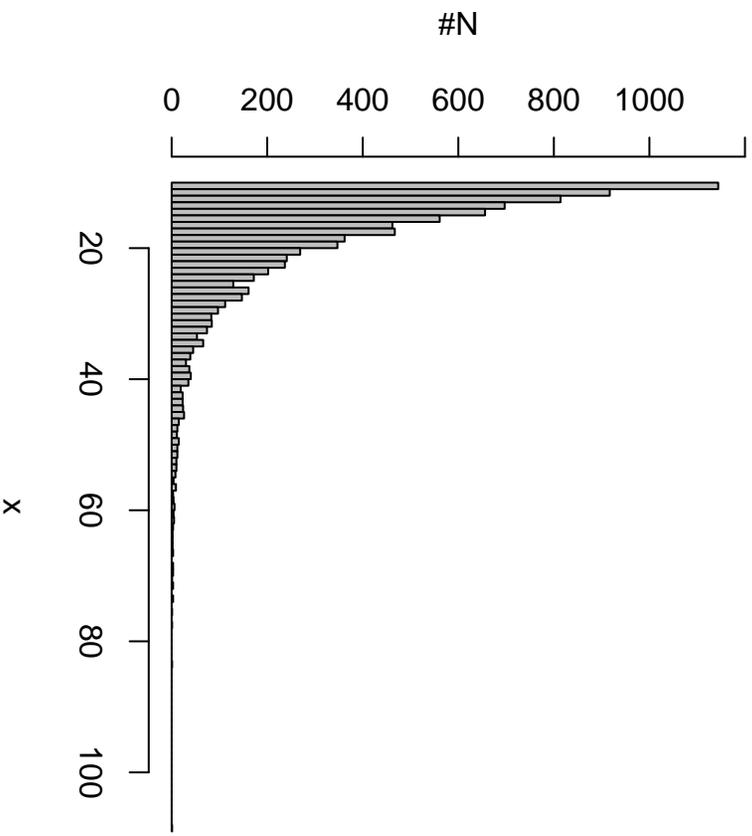
\end{document}